\documentclass[manuscript,screen]{acmart}
\AtBeginDocument{%
  }

\setcopyright{rightsretained}
\copyrightyear{2025}
\acmYear{2025}
\acmDOI{XXXXXXX.XXXXXXX}
\acmConference[CHI '25]{CHI '25 Workshop 'Resisting AI Solutionism'}{June 03--05,
  2018}{Woodstock, NY}
\acmISBN{978-1-4503-XXXX-X/2018/06}

\begin{document}

\title{Emphasizing Deliberation and Critical Thinking in an AI Hype World}

\author{Katja Rogers}
\email{k.s.rogers@uva.nl}
\orcid{0000-0002-5958-3576}
\affiliation{%
  \institution{University of Amsterdam}
  \city{Amsterdam}
  \country{Netherlands}
}

\renewcommand{\shortauthors}{Rogers}

\begin{abstract}
AI solutionism is accelerated and substantiated by hype and HCI's elevation of novelty. Banning or abandoning technology is unlikely to work and probably not beneficial on the whole either---but slow(er), deliberate use together with conscientious, critical engagement and non-engagement may help us navigate a post-AI hype world while contributing to a solid knowledge foundation and reducing harmful impacts in education and research. 
\end{abstract}

\ccsdesc[500]{Computing methodologies~Artificial intelligence}
\ccsdesc[500]{Human-centered computing~Human computer interaction (HCI)}

\keywords{artificial intelligence, hype, novelty, critical thinking}

\received{27 February 2025}

\maketitle

\section{The Hype}
The hype to use artificial intelligence (AI) as a solution for everything doesn't need much belabouring; its existence is clear and extensive. We see this surrounding AI and especially generative AI (genAI) and large-language models (LLMs) in the media \cite{vrabideman2024promising}, in industry \cite{davenport2023all}, and in research \cite{pang2025understanding,liang2024mapping}.
Nature articles discuss which AI tools to use for research tasks like editing manuscripts or grant proposals or brainstorming research ideas and hypotheses~\cite{gibney2025what}.
Researchers in human-computer interaction (HCI) suggest using LLMs to identify stakeholders and their characteristics, to create personas and scenarios, generate interview questions, and possibly even simulate user responses in lieu of real human participants \cite{schmidt2024simulating}---explored more in-depth though somewhat tongue-in-cheek as a way to \textit{"replace
human analysis and then to dispense with human-produced text"} by \citet{byun2023dispensing}.  
Viewed from a more localized, personal perspective: as an HCI researcher in a small, young HCI research group housed within an established AI-focused institute, the hype for AI solutionism \cite{reyes2025resisting} is similarly pervasive and non-abating. It increasingly manifests as pressure to accept AI as an acceptable and/or preferred tool in any shape or context. The teaching centre gives talks about including it in education, colleagues recommend its use for writing grant applications, PhD students and fellow researchers say they use it to write those sections of their work that they do not care about so much, and so on.



Of course, hype in academia is not a new thing. Academic publications have long used hype-based or hype-conveying language (words like "novel" or "innovative" \cite{millar2020hype}) to promote their work and angle for paper acceptance. Papers have multiple purposes, and conveying information is only one of them, even in evidence-based domains like medicine:
\begin{quote}
    \textit{If the purpose of medical research papers were simply to inform, they could largely be reduced to spreadsheets of raw and processed data. However, we would suggest that the narrative form of the medical report persists because of the role of the researcher in deliberately convincing their audience.} \hfill   -- \citet{millar2020hype}
\end{quote}

\begin{figure}
    \centering 
    \includegraphics[trim={0 6cm 0 0},clip, width=0.3\linewidth]{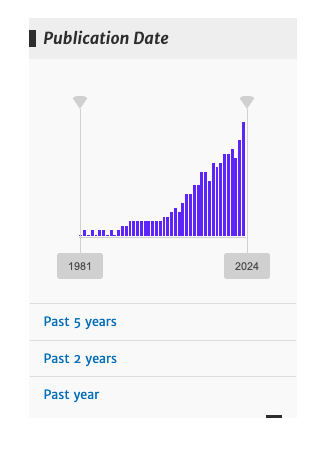}
    ~
    \includegraphics[trim={0 4.9cm 0 0.5cm},clip, width=0.29\linewidth]{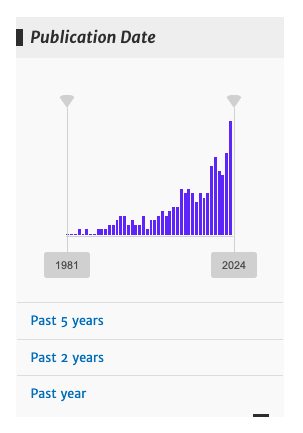}
    \caption{Left: 21,470 results for "novel" in ACM library filtered to CHI proceedings. Right: In comparison, searching for "human-computer interaction" leads to 16,390 results.}
        \label{fig:novelty-in-hci}
\end{figure}

Some degree of hype and novelty may simply be necessary for academic publications and pushing forward new ideas. 
Anecdotally, at least, HCI seems to put a lot of weight on papers' "framing" of the research gap, which is
at times related to whether a paper's contribution is viewed as "novel" enough. This focus on novelty seems also evident in the papers that are published ("novel" is referenced in the CHI proceedings more often than "human-computer interaction", see \autoref{fig:novelty-in-hci}). It is certainly true that innovation and shared excitement can be positive drivers of progress and something that vitalizes academic fields with new ideas. From that perspective, standards and norms for peer review quality---when rigidly applied---might sometimes be viewed as "a challenge for researchers working on pioneering or unconventional ideas" \cite{kaltenhauser2025chi}. Novelty as a subjective criteria is also being explored systematically as a conceptual problem in the design of interactive technology, with a goal of \textit{"less intuitively constructed [...] judgments of potential knowledge contributions"}~\cite{wiberg2014what}. With novelty a favourite of HCI, the AI hype being taken up by HCI is unsurprising, and---in a field focusing on technical innovation---crucial for both HCI as a field and the development and uptake of AI. 

Hype is not a new thing for AI; since its inception, the field has undergone a \textit{"repeating cycle of bubbles and crashes"} also called AI summer and winter \cite{mitchell2019artificial}. 
There are good reasons for excitement---there certainly is substantial potential benefit in time savings resulting from automated or semi-automated approaches to some activities. Using computational tools to correct one's grammar or tone can have undeniable positive impact towards making spaces more inclusive to researchers for whom English is not their first language, or have dyslexia or similar. Automation through AI can be used for social good \cite{hsu2022empowering, tomasev2020ai}. These benefits and others should not be ignored and from a practical / Pandora's box standpoint also likely \emph{cannot} be abandoned at this point.

\section{Resistance through Deliberation, and Critical Thinking Skills}
Generally, technologies have mostly been abandoned when better technology is developed to replace it (e.g., old physical video media formats like VHS and BetaMax). The Luddite revolution against novel industrial weaving machinery did not work out well for them~\cite{skidelsky2017death}. Similarly, I don't expect that regulations to ban AI technologies, or LLMs specifically, are desirable or that it could succeed. 
Yet there are real harms, ongoing and ahead, for example unemployment and job losses, or real-world outcomes resulting from algorithmic embedded bias. According to a recent literature review by \citet{pang2025understanding}, HCI research has already begun exploring \textit{"economic harms, representational harms, misinformation harms, malicious use, hate speech, and environmental harms"} resulting from LLMs.

AI technologies are getting better and better, but I have not yet seen it capable of mastering or even successfully masquerading what I would consider high-level critical thinking skills, and I remain skeptical that it will manage this in the future---though some are hopeful that this is just a matter of time. 
I subscribe to the idea that human judgement and expertise remain a critical component in human-AI interaction \cite{natarajan2024human}. In my view, two key risks arise from abandoning this component: 

\textbf{1)} As someone who teaches, I am worried that \textbf{early reliance on AI could stymie critical thinking}, particularly in students' ability to apply it to their engagement with AI and other systems. Worryingly, a paper at this year's CHI by \citet{lee2025impact} suggests that higher confidence in genAI correlates with less critical thinking. 

\textbf{2)}  I am also worried that relying on AI to produce new content---trained on our current and past state of data---could pose a \textbf{hindrance to efforts of improving the status quo} in the future, both in terms of knowledge generation (e.g., AI in qualitative analysis and knowledge synthesis) and in the values and biases that may be promoted. As just one example for the former, there are several ways in which our reporting of systematic reviews in HCI could be improved~\cite{rogers2025umbrella}, yet if LLMs suggest content based on the current state of things, then poor reporting habits may be amplified and efforts to improve research and reporting practices may become even harder to improve \cite{rogers2025shiny}.

Resistance to AI through  lobbying for regulation and advocating for careful deliberation of trade-offs before its use is necessary, especially given the environmental impacts of its training and deployment \cite{wu2022sustainable}, and the potential social impacts we are only beginning to understand and experience \cite{mantelero2022beyond, whittlestone2022ai}. 
Calls for greater consideration in AI ethics---are the impacts worth using this technology for this purpose?---reflect similar thoughts raised on how we behave about conference travel \cite{jacques2020chi}---another thing we occasionally worry about but then largely ignore (when feasible) because of the strong incentives and benefits attached to in-person networking and conference attendance. In that vein:  

\begin{quote}
    \textit{The purpose of this work is not to guilt-trip or moralise. To be explicit, the message is not “don’t travel.” Rather, these findings should prompt us to ask ourselves, both as individuals and a community, when, where, and why?} \hfill -- \citet{jacques2020chi}
\end{quote}

I firmly believe that a similar approach---i.e., careful weighing of when to use or not use AI, in what ways, and why---needs to be applied to AI solutionism \cite{reyes2025resisting} as well. Pushing back more firmly against reaching for and rewarding quantity and speed in research output is part of that \cite{mhaidli2024overworking}.
Instead, focusing on quality and future quality improvements should be the way forward. 
As strategies, I see the following: 
\textbf{1)} We should \textbf{think before we use or design AI}. For that matter, we should think before we design or deploy technology of any kind \cite{baumer2011when}. 
\textbf{2)} If AI of some kind saves us time, we should consider using the freed time to create social or interpersonal good. If we must use it for work, we should at least \textbf{use freed time to improve the quality and meaningfulness of our work}, instead of creating just more research output. 
\textbf{3)} Exchange ideas on how to \textbf{(re)-design courses to ensure that learning objectives include and assess critical thinking skills}. For example, I am currently considering new (as well as in some cases more traditional) exam formats and classroom activities to avoid grading ChatGPT output. 

Overall, for me, resisting AI solutionism means resistance through slow(er) research, careful deliberation, and conscientious, critical engagement with AI---along with personal commitment and community advocacy and incentives to consider \emph{non-}engagement---in the design of applications, HCI and AI education, and our research and work practices.  


\begin{acks}
Thank you to Alex Fleck and Siân Brooke for helpful conversations and comments on an earlier draft. 
\end{acks}

\bibliographystyle{ACM-Reference-Format}
\bibliography{sample-base}

\appendix

\end{document}